\documentclass[aps,twocolumn,showpacs,floatfix]{revtex4}
\usepackage{amsmath,amssymb,isolatin1,graphicx,epsfig,times}

\begin{document}

\title{Dynamics of end-linked star polymer structures}
\author{C. Satmarel}
\author{C. von Ferber}
\author{ A. Blumen}
\email{blumen@physik.uni-freiburg.de}

\affiliation{ Theoretische Polymerphysik, Universit\"at Freiburg, Hermann-Herder Str. 3, D-79104 Freiburg, Germany }

\date{\today}

\begin{abstract}
In this work we focus on the dynamics of macromolecular networks formed by
end-linking identical polymer stars. The resulting macromolecular network can then be
viewed as consisting of spacers which connect branching
points (the cores of the stars). We succeed in analyzing exactly, in the framework of the generalized Gaussian model, the eigenvalue spectrum of such networks. As applications we focus on several topologies, such as regular networks and dendrimers; furthermore, we compare the results to those found for regular hyperbranched structures. In so doing, we also consider situations in which the beads of the cores differ from the beads of the spacers. The analytical procedure which we use involves an exact real-space renormalization, which allows to relate the star-network to a (much simpler) network, in which each star is reduced to its core. It turns out that the eigenvalue spectrum of the star-polymer structure consists of two parts: One follows in terms of polynomial equations from the relaxation spectrum of the corresponding renormalized structure, while the second part involves the motion of the spacer chains themselves. Finally, we show exemplarily the situation for copolymeric dendrimers, calculate their spectra, and from them their storage and the loss moduli.
\end{abstract}
\maketitle

\section{Introduction}

A series of recent works focuses on the synthesis of macromolecular networks
by end-linking star polymers \cite{dushek,gasilova,sukumar,prochazka,burchard,erwan2001}. Now, previous work \cite{vicsek} has considered hyperbranched polymers, which were iteratively built from rather
small units: there the use of regular repeated units
allowed a complete, analytical determination of the eigenfrequencies. Here we extend this
work by focusing on structures built from stars. Using stars as building blocks allows to vary both $f$, the functionality and also $k$, the numbers of monomers along the arms. The stars will be envisaged as being end-linked in such a way that two end monomers
belonging to two different stars get connected by an additional link. In
this way we obtain a system of branching points connected by spacers; this
system of branching points may, on the other hand, be rather arbitrary, and
may by itself form a complex network such as a regular dendrimer \cite{dushek}, a branched polymer \cite{burchard,vicsek} or a branched copolymeric structure \cite{vamvakaki}.

The simplest method to determine the dynamics of such networks is to treat them as  generalized Gaussian structures (GGS) \cite{sommer_blumen,doi_edwards,rama,gurtovenko_blumen}: the polymer is viewed as consisting of beads that are linked by harmonic springs. In this way, the classical Rouse-model gets adapted to general topologies. As it is well-known, the Rouse-model describes a chain molecule immersed in a viscous solvent, so that
for each monomer the damping due to the solvent and the harmonic potential due to the other monomers are
accounted for. On the other hand, the Rouse-model neglects hydrodynamic interactions, inertia terms,
the excluded volume and entanglement effects. Now, the
limitations of the Rouse model are well understood; although not very
realistic, Rouse-models and their extensions are certainly very useful in
unraveling the dynamics of complex structures. \cite{doi_edwards,gurtovenko_blumen} 

In the study of networks built from stars we will take all the springs to be
equal, but we will differentiate between the monomers at branching points and
the monomers along the spacers; we will let the two kinds of monomers have
different mobilities i.e. different friction coefficients. The reason for this
are twofold: first, in many cases stars are prepared by having a central monomer which differs from the other monomers; second, mathematically this generalization does not significantly burden the formalism.
This procedure, on the other hand, allows us to discuss easily the case of
copolymers. Now, as was shown in former work \cite{sgb1,sgb2}, the relaxation of
copolymeric structures containing units of different mobility often differs
qualitatively from the relaxation of similar homopolymeric structures, in which all the beads are the same \cite{sgb1,sgb2}.

In what the relaxation dynamics of the networks is concerned, we focus on
the complex dynamical shear modulus, $G^{\ast}(\omega),$ or in more
conventional forms, on its real and imaginary parts, which are the dynamic storage and the loss modulus, $G^{\prime}(\omega)$ and $G^{\prime\prime}(\omega)$, respectively. The advantage of looking at these quantities is, first, that they are widely measured, and, second, that there are intimately related to other dynamic quantities, such as the dielectric and the magnetic relaxation. Moreover, in the evaluation of $G^{\prime}(\omega)$ and $G^{\prime\prime}(\omega)$ only the eigenvalues of the underlying connectivity matrix enter, but not the eigenvectors \cite{doi_edwards,rama,gurtovenko_blumen}. This fact considerably simplifies the mathematical treatment, by allowing us to focus on the eigenvalues. As we proceed to show we can relate the eigenvalues of our system of end-linked stars to the corresponding underlying spacer-free structure.

The basic realization here is that the problem considered here admits an {\it exact} real-space renormalization. This may appear surprising after the efforts displayed up-to-now \cite{kloczkowski,grassley}, which lead only to {\it approximate} results. The interesting feature is that the previous works used models in which some beads were assumed to be fixed \cite{kloczkowski,grassley}. Surprisingly, it is just this assumption which {\it complicates} the analytical solution; by allowing all beads to move freely the problem simplifies (of course, some normal modes leave groups of beads immobile, but this is a result, and not an a priori assumption). The situation here parallels that found in Ref. \cite{vicsek}, where regular hyperbranched polymers were investigated; the exact results in \cite{vicsek} followed earlier works, which were unable to solve the problem exactly, again since they took the positions of the peripheral beads to be fixed. 

The paper is structured as follows: In the next section we present the model considered and recall the forms of $G^{\prime}(\omega)$ and of $G^{\prime\prime}(\omega)$ in the GGS-formalism. In Sec. 3 we explain our renormalization procedure, which allows us to relate analytically the spectra of the networks under investigation to those of simpler structures, in which the spacers and the dangling chains are eliminated. In Sec. 4 we display the use of our method in several cases and focus on star polymers, on a cube polymer and on topologically periodic networks. Dendritic structures made out of copolymeric stars are analyzed in Sec. 5, where we present numerical results for $G^{\prime}(\omega)$ and $G^{\prime\prime}(\omega)$. Our conclusions are summarized in Sec. 6, while the mathematical details are collected in four Appendices.

\section{The Model}

\begin{figure}
\begin{center}
\epsfig{file=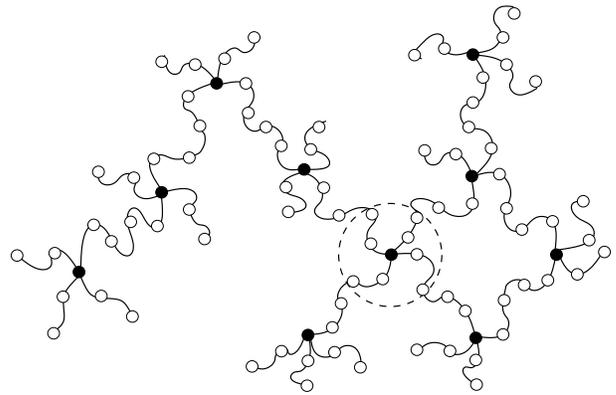,width=8cm,clip}
\end{center}
\caption{Network consisting of end-linked polymer stars, of which one is highlighted by a circle. Here $N=10$,$f=4$, and $k=2$, so that $N_{full}=90$. Note that the structure is very flexible so that in many realizations it will appear considerably less ordered than here.}
\label{linked_stars}
\end{figure}

In the following we consider a system of $N$ end-linked polymer stars. Each star consists of a core (a central $f$-coordinated bead) from which emerge $f$ chains of $k$ beads each. The central bead may differ (chemically and functionally) from the other beads, a fact which we take into account by assigning it a different friction coefficient. In Fig. \ref{linked_stars} we show such a network formed from end-linked stars. Corresponding to the coloring used in Fig. \ref{linked_stars}, we call in the following the core bead ``black'' and the chain beads ``white''. Note that in Fig. \ref{linked_stars} the stars are linked at their ends, through the means of $l$ additional bonds, to form a single, connected cluster. The linking topologies which we consider are rather arbitrary; the restriction that we impose is that the cores (branching points) form a simple graph structure. By this we forbid links between chains belonging to the same star as well as multiple links between two stars.

Now we model the system of Fig. \ref{linked_stars} in the framework of generalized Gaussian structures (GGS), which are an extension of the Rouse-model for Gaussian chains \cite{sommer_blumen,rama,gurtovenko_blumen}. We focus on networks of beads connected by harmonic springs with elastic constant $K$, and we take the networks to be immersed in a viscous solvent, where the inertial forces can be neglected. The corresponding Langevin equations of motion for the system have the form:
\begin{equation}
\label{langevin_lambda}
\zeta_j\frac{\partial{\bf r}_j(t)}{\partial t}-K\sum_{\alpha =1}^{n_j}\Big[{\bf r}_{i_\alpha}(t)-{\bf r}_j(t)\Big]={\bf F}({\bf r}_j,t)+{\bf f}_j(t).
\end{equation}
Equation (\ref{langevin_lambda}) expresses the fact that for each bead, say bead $j$ at position ${\bf r}_j(t)$, the friction force and the elastic forces acting on it exactly balance the external and the fluctuating forces. In Eq. (\ref{langevin_lambda}), the ${\bf r}_{i_\alpha}(t)$ with $\alpha=1,\ldots ,n_j$ denote the positions of the $n_j$ neighbors of bead $j$, $\zeta_j$ is its friction coefficient, and ${\bf f}_j(t)$ and ${\bf F}({\bf r}_j,t)$ are the respective random and external forces; the fluctuating forces are taken to be Gaussian, with $\langle f_{j,a}(t)f_{i,b}(t')\rangle=2k_BT\zeta_j\delta_{ij}\delta_{ab}\delta(t-t')$. Moreover, we take the friction coefficients of the white beads to be $\zeta_j=\zeta_0$ and of the black beads to be $\zeta_j=\sigma\zeta_0$.

In the GGS picture $G'(\omega)$, the storage and $G''(\omega)$, the loss  modulus are given in terms of the eigenvalues $\lambda_i$ of the homogeneous version of Eq. (\ref{langevin_lambda}) (where the right-hand-side is put equal to zero). One has namely, in a reduced form \cite{rama,gurtovenko_blumen,gurt02} 
\begin{equation}
\label{gprim}
G'(\bar\omega)=\frac{1}{N_{full}}\sum_{m=2}^{N_{full}}\frac{\bar\omega^2}{\bar\omega^2+4\lambda_m^2}
\end{equation}
and
\begin{equation}
\label{gdprim}
G''(\bar\omega)=\frac{1}{N_{full}}\sum_{m=2}^{N_{full}}\frac{2\bar\omega\lambda_m}{\bar\omega^2+4\lambda_m}.
\end{equation}
In Eqs. (\ref{gprim}) and (\ref{gdprim}) the reduced frequency $\bar\omega=\omega/\tau_0$ is given in terms of the fundamental time $\tau_0=\zeta_0/K$, $N_{full}$ is the total number of beads and the $\lambda_m$ are the $N_{full}-1$ non vanishing eigenvalues of the system, where 
\begin{equation}
\label{1a}
N_{full}=N(fk+1),
\end{equation}
given that each star is composed of a core (black bead) and of $fk$ white beads. As shown previously, \cite{sgb1,sgb2} the general forms of $G'(\omega)$ and $G''(\omega)$, well-known from the study of homopolymers, also hold for heteropolymeric networks, such as the ones considered here.

To determine the eigenvalues of the left-hand-side of  Eq. (\ref{langevin_lambda}) we focus on its normal modes, which have the form ${\bf P}_j^{(m)}(t)=\Phi_j^{(m)}\exp{(-\lambda_m t/\tau_0)}$. Inserting them in the left-hand-side of Eq. (\ref{langevin_lambda}) leads to the following system of equations: 
\begin{equation}
\label{n_j_coord}
(n_j-\frac{\zeta_j}{\zeta_0}\lambda_m)\Phi_j^{(m)}=\sum_{\alpha=1}^{n_j}\Phi_{i_\alpha}^{(m)},~~~\textrm{with}~1\le j\le N_{full}.
\end{equation}
With $N_{full}$ being the total number of beads in the system, this represents $N_{full}$ equations. As is well-known from the general theory, the linear system given by Eq. (\ref{n_j_coord}) leads to $N_{full}-1$ positive eigenvalues, $\lambda_m>0$, and to a single vanishing eigenvalue, $\lambda_1=0$. 

In the following we look separately at the system of equations belonging to each $m$ and for ease of notation drop the index $m$. For the ${\it f}$-coordinated core (black bead) of each star, Eq. (\ref{n_j_coord}) reads:
\begin{equation}
\label{f_coord}
(f-\sigma\lambda)\Phi_j=\sum_{\alpha=1}^{f}\Phi_{i_{\alpha}}.
\end{equation}
For the white beads of the spacers (these are doubly coordinated), the corresponding equation is
\begin{equation}
\label{2_coord}
(2-\lambda)\Phi_{i+1}=\Phi_{i}+\Phi_{i+2},
\end{equation}
where we numbered the beads consecutively along the spacer. Again from Eq. (\ref{n_j_coord}) we have for every peripheral bead $k$ (which is ${\it singly}$-coordinated)
\begin{equation}
\label{1_coord}
(1-\lambda)\Phi_k=\Phi_{k-1}.
\end{equation}

Remarkably, the system of Eqs. (\ref{f_coord}), (\ref{2_coord}) and (\ref{1_coord}) can be simplified analytically for any arrangement like the one in Fig. \ref{linked_stars}. This fact extends to stars the exact mathematical treatment developed for small building blocks, used in the study of regular hyperbranched structures.

\section{Renormalization procedure}

The following section is devoted to the simplification of the system of Eqs.
(\ref{f_coord}), (\ref{2_coord}) and (\ref{1_coord}). The basic idea is that one has an analytical procedure at
one's disposal, by which one can "delete" most of the beads of the original
system, namely those on the spacers and on the dangling chains; in this way one
is led to a smaller, reduced system, consisting of the (directly-connected)
black beads only. The procedure is exact and we will develop it in the main
body of this section. The method parallels that of determining the
eigenvalues of regular hyperbranched structures, a problem treated in Refs. \cite{wedges,cai,kloczkowski,grassley}, but it is more extensive. Moreover, it will turn out that the eigenvalues of the original system
fall naturally into two classes: The first class is the one related to
non-vanishing eigenvalues of the reduced black system, while for the second
class the corresponding eigenvalue of the reduced system is zero. As a
consequence, we follow a step-by-step development and focus first on the spacers of a system such as displayed in Fig. \ref{linked_stars}, i.e. on the chains of white beads.

\subsection{General framework}

Consider now Fig. \ref{linked_stars}, in which for each pair of linked stars the core black beads are connected by a chain of $s=2k$ white beads, and in which each dangling chain consists of $k$ white beads. We consider first an arbitrary chain of white beads and number them sequentially denoting the starting black bead by ``0''. In this way the amplitudes (eigenmode components) of the white beads get denoted by $\Phi_1, \Phi_2,\ldots$, while that of the black bead gets denoted by $\Phi_0$. For a dangling chain the last amplitude is $\Phi_k$, while for a spacer it is $\Phi_s$, with $\Phi_{s+1}$ denoting the amplitude of the next black bead.  

Now, all the white beads fulfill Eq. (\ref{2_coord}). Introducing the notation  
\begin{equation}
\label{2x}
2x=2-\lambda,
\end{equation} 
Eq. (\ref{2_coord}) reads exemplarily for the white bead $i+1$: 
\begin{equation}
\label{text_2}
\Phi_{i}=2x\Phi_{i+1}-\Phi_{i+2}.
\end{equation}
The following considerations are significantly simplified by making use of the Chebyshev polynomials $U_n(x)$ of the second kind, see Eq. (22.2.5) of Ref. \cite{Abramovitz}. The first two Chebyshev polynomials read: $U_{0}(x)=1$ and $U_1(x)=2x$, Table 22.5 of Ref. \cite{Abramovitz}, where they are standardized by requiring that $U_n(1)=n+1$. Furthermore, the $U_n(x)$ obey the recursion relation:
\begin{equation}
\label{text_3}
U_n(x)=2xU_{n-1}(x)-U_{n-2}(x),
\end{equation}
see Eq. (22.7.5) of Ref. \cite{Abramovitz}. In terms of trigonometric functions one also has Eq. (22.3.16) of Ref. \cite{Abramovitz}:
\begin{equation}
\label{text_9}
U_n(\cos \varphi)=\frac{\sin(n+1)\varphi}{\sin \varphi},
\end{equation}
where for $\varphi\in\mathbb{R}$ one has $-1\le x\le 1$. Of course, as polynomials the $U_{f,s}(x)$ are defined over $\mathbb{R}$, which means that $\varphi$ in Eq. (\ref{text_9}) is allowed to be complex. 

We now proceed to show by induction that the amplitudes of each pair $(n,n+1)$ of white beads along the chain are related to that of the black bead through
\begin{equation}
\label{text_4}
\Phi_0=U_n\Phi_n-U_{n-1}\Phi_{n+1},
\end{equation}
where here and in the following $U_n(x)\equiv U_n$ is implied, if not indicated otherwise.

Now, for $n=1$ Eq. (\ref{text_4}) reproduces Eq. (\ref{text_2}) for $i=0$. Let us therefore prove the induction step. Inserting Eq. (\ref{text_2}) into Eq. (\ref{text_4}) for $n=i$ we find
\begin{eqnarray}
\nonumber
\Phi_0 & = & U_n(2x\Phi_{n+1}-\Phi_{n+2})-U_{n-1}\Phi_{n+1}= \\
\nonumber
& = & (2xU_n-U_{n-1})\Phi_{n+1}-U_n\Phi_{n+2} \\
\label{text_5}
& = & U_{n+1}\Phi_{n+1}-U_{n}\Phi_{n+2},
\end{eqnarray}
where the recursion relation, Eq. (\ref{text_3}), was used in the last line. This proves that if Eq. (\ref{text_4}) is valid for $n$, it is also valid for $n+1$, which completes the induction step.  Now, no particular use was made of the fact that $\Phi_0$ stands for the black bead. We can as well start with bead $n+1$ and work our way back to bead $0$. In this way we obtain:
\begin{equation}
\label{text_6}
\Phi_{n+1}=U_n\Phi_1-U_{n-1}\Phi_0.
\end{equation}

A fundamental, special case of this equation relates the amplitudes of neighboring black beads connected by a spacer:
\begin{equation}
\label{13a}
\Phi_{s+1}=U_s\Phi_1-U_{s-1}\Phi_0.
\end{equation}
Let us now turn to a dangling chain of $k$ beads, whose amplitudes are $\Phi_1,\ldots ,\Phi_k$. Now, from Eq. (\ref{1_coord}) one has
\begin{equation}
\label{text_6b}
(2x-1)\Phi_k=(1-\lambda)\Phi_k=\Phi_{k-1}.
\end{equation}
 Moreover, from Eq. (\ref{text_6}) we have
\begin{equation}
\label{1}
\Phi_k  =  U_{k-1}\Phi_1-U_{k-2}\Phi_0
\end{equation}
and
\begin{equation}
\label{2}
\Phi_{k-1}  =  U_{k-2}\Phi_1-U_{k-3}\Phi_0.
\end{equation}
Inserting these into Eq. (\ref{text_6b}) we obtain:
\begin{equation}
(2x-1)(U_{k-1}\Phi_1-U_{k-2}\Phi_0)  =  U_{k-2}\Phi_1-U_{k-3}\Phi_0 
\end{equation}
and hence:
\begin{equation}
[(2x-1)U_{k-1}-U_{k-2}]\Phi_1  =  [(2x-1)U_{k-2}-U_{k-3}]\Phi_0.
\end{equation}
Making now use of Eq. (\ref{text_3}) it follows
\begin{equation}
\label{important}
(U_k-U_{k-1})\Phi_1=(U_{k-1}-U_{k-2})\Phi_0.
\end{equation} 
Multiplying this equation with $(U_k~+~U_{k-1})$ and making use of Eqs. (\ref{old22}) and (\ref{old21}) of Appendix A leads to our second fundamental relation:
\begin{equation}
\label{22}
U_{2k}\Phi_1=(U_{2k-1}+1)\Phi_0.
\end{equation}

Now, let us consider a general black bead of amplitude $\Phi_0$
that is connected to $f_1$ neighboring black beads (with amplitudes
$\Phi_{s+1}^{(1)},\ldots ,\Phi_{s+1}^{(f_1)}$) via chains of $s=2k$ beads
and that has attached to it $f_2$ dangling chains of $k$ beads each, see Fig. \ref{linked_stars}. Then the black bead is $f$-coordinated, $f=f_1+f_2$. Equation (\ref{f_coord}) corresponding to this situation reads:
\begin{equation}
\label{text_10}
(f_1+f_2-\sigma\lambda)\Phi_0=\sum_{a=1}^{f_1}\Phi_1^{(a)}+\sum_{b=1}^{f_2}\Phi_1^{(f_1+b)},
\end{equation}
where, in an obvious way, $\Phi_1^{(a)}$ denotes the amplitude of the first bead of chain $a$. We now multiply Eq. (\ref{text_10}) by $U_s$ and insert the fundamental Eqs. (\ref{13a}) and (\ref{22}) into it, obtaining
\begin{eqnarray}
\nonumber
(f_1+f_2-\sigma\lambda)U_s\Phi_0 & = & \sum_{a=1}^{f_1}(\Phi_{s+1}^{(a)}+U_{s-1}\Phi_0) \\
& + & f_2(U_{s-1}+1)\Phi_0.
\label{text_11}
\end{eqnarray}
This can be rewritten as
\begin{equation}
\label{text_12}
[f_1-P_{f,s}(\lambda)]\Phi_0=\sum_{a=1}^{f_1}\Phi_{s+1}^{(a)},
\end{equation}
with
\begin{equation}
\label{text_13}
P_{f,s}(\lambda)=f+fU_{s-1}-(f-\sigma\lambda)U_s.
\end{equation}

Here we recall that $U_s$ is a polynomial of order $s$ in the variable $x=1-\lambda /2$ (and analogously for $U_{s-1}$), so that $P_{f,s}(\lambda)$ is a polynomial of order $s+1$ in $\lambda$. Equation (\ref{text_12}) is the fundamental result for our model and it relates directly neighboring black beads; the white beads have been eliminated from it. We discuss it extensively in the next subsection. Evidently, in general one assumes that $U_s\ne 0$, otherwise the multiplications in Eqs. (\ref{22}) and (\ref{text_11}) lead to a loss of information. In the following we will take particular care of the cases where $U_s=0$.

\subsection{Determination of the eigenvalue spectrum}
Let us interpret this remarkable result, Eq. (\ref{text_12}). This equation relates the amplitude of every black bead to the amplitudes of its neighboring black beads only; all reference to the white beads on the spacers and on the dangling chains has been eliminated. We denote the eigenvalues of the reduced black bead system by $\lambda_b$ and have, rewriting Eqs. (\ref{text_12}) and (\ref{text_13}):
\begin{equation}
\label{text_15}
(f_1-\lambda_b)\Phi_0=\sum_{a=1}^{f_1}\Phi_{s+1}^{(a)} 
\end{equation}
with
\begin{equation}
\label{text_16}
P_{f,s}(\lambda)=\lambda_b.
\end{equation}
Remarkably thus (as in the case of the iterative procedure used for fractal nets \cite{vicsek,rammal}) each of the $N-1$ nonzero eigenvalues $\lambda_b\neq 0$ of the black subsystem leads via Eq. (\ref{text_16}) to $s+1=2k+1$ eigenvalues of the full network (white and black beads). These eigenvalues correspond to eigenmodes of the system that involve the concomitant motion of white and black beads.

In this way we have determined, given that our cluster is composed of $N$
stars, a total of 
\begin{equation}
\label{29a}
N_P=(N-1)(2k+1)
\end{equation}
eigenvalues. We recall that the whole system must have $N_{full}=N(fk+1)$ eigenvalues, see Eq. (\ref{1a}). Evidently, the remaining (in general, many) eigenvalues are connected to the zero eigenvalue, $\lambda_{b}=0$, of the black subsystem. 

 The question arises to determine all the modes of the full system of white and black beads which lead to $\lambda_b=0$. Now in this case, as is known from
general considerations \cite{doi_edwards,rama}, the amplitude of all black beads is the same. This fact is also reflected by Eq.(\ref{text_15}), which clearly admits as a solution
\begin{equation}
\label{text_17}
\Phi_{s+1}^{(a)}=\Phi_0 \mbox{,\hspace{0.3cm}} \textrm{for}\hspace{0.2cm} a=1,\ldots,f_1.
\end{equation}
The simplest mode with $\lambda_b=0$ and $\Phi_0\ne 0$ is the translation of {\it all} the beads in parallel; it corresponds to the eigenvalue $\lambda_1=0$ {\it of the full system}. However, $\lambda_b=0$ is also related to other modes in which the black beads move in parallel, and gives rise to additional eigenvalues of the full system.

We now turn to consider the black subsystem. The simplest situation occurs
when all black beads are at rest. In order to avoid a trivial solution one has
imperatively to ask that (at least) one of the white beads neighboring a black
one has as amplitude $\Phi_{1}\neq0$. Since the amplitude of its neighboring
black bead vanishes, $\Phi_{0}=0$, this implies, as a necessary condition for
non-triviality, that $U_{s}=U_{2k}=0$ holds both for the spacers, Eq. (\ref{13a}) (since also $\Phi_{s+1}=0$),
as well as for the dangling chains, Eq. (\ref{22}). We are hence in both cases having
$U_{s}=0$; recalling the form of $U_s$ given in Eq. (\ref{old21}) of Appendix A this means:
\begin{equation}
\label{34}
U_{s}=U_{2k}=(U_{k}+U_{k-1})(U_{k}-U_{k-1})=0.
\end{equation}
Moreover, consideration of Eq. (\ref{important}) shows that for dangling chains
one has to sharpen the condition and to ask even that %
\begin{equation}
\label{35}
U_{k}-U_{k-1}=0.
\end{equation}
 
From Eq. (\ref{34}) we thus obtain two kinds of normal modes. One set of them
obeys Eq. (\ref{35}) and may involve both dangling chains and also spacers,
whereas the second set of normal modes obeys
\begin{equation}
\label{36}
U_{k}+U_{k-1}=0,
\end{equation}
and can be fulfilled only by modes localized on the spacers exclusively.

Now the roots of $U_s=U_{2k}=0$ are, using Eq. (\ref{text_9}), given by $\sin[(2k+1)\varphi]=0$ under the condition that $\sin\varphi\ne 0$. Hence they have the form
\begin{equation}
\label{40}
\varphi_r=\frac{r\pi}{2k+1},~~~~~~~\textrm{with}~~~r=1,\ldots,2k.
\end{equation}
From this relation the roots $x_r$ follow from $x_r=\cos\varphi_r$. Given the form of Eq. (\ref{40}) it is clear that none of them equals unity, $x_r\ne 1$, and thus none of them leads to $\lambda_r=0$, see Eq. (\ref{2x}).

Moreover, not all of these $x_r$ (as stressed above) may appear when white beads of dangling chains move. For these Eq. (\ref{35}), namely
\begin{equation}
\label{text_18a}
U_k(\cos\varphi) - U_{k-1}(\cos\varphi)=0
\end{equation}
must hold. From Eq. (\ref{text_9}) the roots of Eq. (\ref{text_18a}) are the roots of $\sin[(k+1)\varphi]-\sin(k\varphi)=0$ for which $\sin\varphi\neq 0$. Since $\sin(\alpha+\beta)-\sin(\alpha-\beta)=2\sin\beta\cos\alpha$, the corresponding roots are:
\begin{equation}
\label{text_18b}
\varphi_m = \frac{2m-1}{2k+1}\pi \mbox{,\hspace{0.3cm}}\hspace{0.5cm} \textrm{with}~~~~~~~ m=1,\ldots,k,
\end{equation}
whereas the modes involving spacers can also have, see Eq. (\ref{40}),
\begin{equation}
\label{loops_b}
\varphi_m=\frac{2m}{2k+1}\pi {,\hspace{0.5cm}} \hspace{0.5cm} \textrm{with}~~~~~~~ m=1,\ldots,k.
\end{equation}
A small remark concerns now the symmetry of the modes considered. From Eq. (\ref{text_4}) we have, since $\Phi_0=0$ that
\begin{equation}
\label{new2}
\Phi_n/U_{n-1}=\Phi_{n+1}/U_n.
\end{equation}
From the point of view of the spacers, this implies that the eigenvalues corresponding to Eq.  (\ref{loops_b}) have eigenmodes which are antisymmetric with respect to the mid-chain beads $(k,k+1)$, whereas the eigenvalues corresponding to Eq. (\ref{text_18b}) have symmetric eigenmodes.

The modes which we have now determined and which correspond to $U_{s}=0$ are
highly degenerate. We display the evaluation of their (corresponding)
multiplicities in Appendix B. The basic idea developed in Appendix B is that,
due to Eq. (\ref{f_coord}), for $\Phi_{0}=0$, a non-vanishing value of $\Phi_{1}^{(a)}$
cannot appear alone, but must be compensated by the motion of (at least)
another bead. Hence, if one starts to look for linearly independent eigenmodes it is reasonable to normalize one of the amplitudes, and set $\Phi_{1}^{(a)}=1$. Moreover, one has to satisfy Eq. (\ref{f_coord}). Again, due to simplicity, one assumes that all the other white neighbors of the given black bead are at rest, except for another one, or, at most two. This leads to having, in an obvious notation $\Phi_{1}^{(b)}=-1$ or $\Phi_1^{(b)}=\Phi_1^{(c)}=-1/2$, respectively. In this way one is able to count all the linearly independent pathways through the full system, in which the signs of the modes are alternating, and which, depending on the eigenvalue considered, may of may not involve dangling chains. 

This procedure is performed in Appendix B, where we show that the multiplicities of the eigenvalues to Eq. (\ref{text_18b}) and to Eq. (\ref{loops_b}) are (quite generally) given by:
\begin{equation}
\label{delta+}
\Delta_+ = l-(N-1) 
\end{equation}
and by
\begin{equation}
\label{delta-}
\Delta_- = (f-1)N-l,
\end{equation}
where the indices indicate that they belong to Eqs. (\ref{36}) and (\ref{35}), respectively. As before, in Eqs. (\ref{delta+}) and (\ref{delta-}) $N$ denotes the number of stars and $l$ the number of additional links connecting them. Then $\Delta_+$ is related to the number of loops in the system. For $\Delta_+=0$ the structure is topologically a tree, no loops exist, and eigenvalues to Eq. (\ref{loops_b}) do not appear in the spectrum. For $\Delta_+=0$, furthermore, $\Delta_-=(f-2)N+1$, and $\Delta_-+1$ is the number of dangling chains. From Eqs. (\ref{delta+}) and (\ref{delta-}) the total number of eigenvalues arising from this class of modes is
\begin{equation}
\label{41a}
N_{\Delta}=k(\Delta_++\Delta_-)=k[(f-2)N+1].
\end{equation}

As a side-remark, we note that in a very special situation (seldom encountered in randomly linked systems), for a particular value of $\lambda_b$ in Eq. (\ref{text_16}) $k$ modes of the black and white system get to be such as to have $U_s=0$. The corresponding modes belong then to both classes, Eq. (\ref{text_16}) and Eq. (\ref{delta-}). (Evidently, they may only be counted once). The situation is discussed in Appendix C.

Finally we have to consider the situation in which all black beads are moving in parallel and the motion of the white beads compensates it. Then usually $U_s\ne 0$ and one is again led to reconsider $P_{f,s}(\lambda)$, but now, since $\lambda_b=0$ one has
\begin{equation}
\label{text_18}
P_{f,s}(\lambda)=0.
\end{equation}
Moreover, using the relation $s=2k$ and Eqs. (\ref{old22}) and (\ref{old21}) of Appendix A with $n=k$, we remark that $P_{f,s}(\lambda)$, Eq. (\ref{text_13}), can be reformulated as
\begin{equation}
\label{text_19a}
P_{f,s}(\lambda)=Q_{f,s}(\lambda)(U_k+U_{k-1}),
\end{equation}
where we set 
\begin{equation}
\label{def_Q}
Q_{f,s}(\lambda)=f(U_{k-1}-U_{k-2})-(f-\sigma\lambda)(U_k-U_{k-1}).
\end{equation}
From Eq. (\ref{text_19a}) and recalling the remarks after Eq. (\ref{text_13}), $P_{f,s}(\lambda)=0$ requires 
\begin{equation}
\label{Q=0}
Q_{f,s}(\lambda)=0,
\end{equation} 
whose modes involve spacers and dangling chains alike. As a side-remark, it is sometimes convenient to formulate $Q_{f,s}(\lambda)$ in an alternate way. From the recursion relation, Eq. (\ref{text_3}), one has 
\begin{equation}
\label{new_number}
(U_{k-1}-U_{k-2})-(U_k-U_{k-1})=2U_{k-1}-2xU_{k+1}=\lambda U_{k-1},
\end{equation}
with which Eq. (\ref{def_Q}) takes the form:
\begin{equation}
\label{Q_simplified}
Q_{f,s}(\lambda)=\lambda[(f-\sigma)U_{k-1}+\sigma U_k].
\end{equation} 
Turning now to the general features of $Q_{f,s}(\lambda)$ we remark that its degree, as given by Eq. (\ref{def_Q}), is $k+1$, and hence we expect in general
\begin{equation}
\label{44a}
N_Q= k+1 
\end{equation}
different eigenvalues from Eq. (\ref{Q=0}). By inspection of Eq. (\ref{Q_simplified}) it is evident that $\lambda_{1}=0$ is indeed a solution of Eq. (\ref{Q=0}). Hence the translation of the full system, which is related to $\lambda_{1}=0$, is included in the set of modes corresponding to Eq. (\ref{Q=0}). This solution, as well as the other $k$ additional ones of Eq. (\ref{Q=0}), is non-degenerate.

\subsection{Classification of eigenvalues and special cases}
Summarizing, we have found a set of equations that fully determine the
eigenvalue spectrum of our system. These are, namely, Eqs. (\ref{text_16}), (\ref{35}), (\ref{36}) and (\ref{Q=0}). Evidently, they
belong to different types of modes, which are clearly linearly independent of
each other. In the first class, given by Eq. (\ref{text_16}), the coordinates of some of
the black beads are non-vanishing and the black beads do not move in parallel. The second
class, given by Eq. (\ref{Q=0}), is that in which all the black beads move in
parallel. The next two classes, exemplified by Eqs. (\ref{35}) and (\ref{36}), are such
that all the black beads are at rest.

Let us check the total number of eigenvalues that we get from these modes. We find from Eqs. (\ref{text_16}), (\ref{41a}) and (\ref{Q=0}) a total of
\begin{eqnarray}
\nonumber
& & N_P+N_Q+N_{\Delta} = \\
\nonumber
& & = (2k+1)(N-1)+(k+1)+k[(f-2)N+1] \\ 
\label{N_full}
& &=  (fk+1)N=N_{full}
\end{eqnarray}
linearly independent eigenmodes, where in the last expression we recalled Eq. (\ref{1a}). By this we have indeed found a complete set of eigenmodes of the full system, leading to the correct, total number of eigenvalues.

To end this section we stress that the determination of the multiplicities Eqs. (\ref{delta-}) and (\ref{delta+}) for the modes given by Eqs. (\ref{35}) and (\ref{36}), i.e. to the relation $U_s=0$, see Eq. (\ref{34}), was very general and depended only on the topology of our network. Eqs. (\ref{delta-}) and (\ref{delta+}) were discussed in the framework of normal modes in which all black beads are at rest; the degeneracy of their eigenvalues is inherent from the network's structure.

Interestingly, in special cases (special types of lattices or parameter values), one may also be led to the relation $U_s=0$, albeit for the other two types of normal modes. One obtains then {\it additional} degeneracies, which are, however, accidental. Thus for lattices without dangling chains and where the black beads build two alternating sublattices (see Appendix C) eigenvalues to $P_{f,s}(\lambda)=\lambda_b=2f$ turn out to be of the type $U_s=0$. Furthermore, the choice $\sigma=f/2$ for the parameter $\sigma$ transforms Eq. (\ref{Q=0}), $Q_{f,s}(\lambda)=0$, using Eq. (\ref{Q_simplified}) to the condition $U_{k-1}+U_k=0$, which automatically implies $U_s=U_{2k}=0$, see Eq. (\ref{old21}). 

\section{Applications}
In this Section we apply our method to several types of lattices made out of
star polymers. In order to exemplify our ideas we start with the simplest
case, namely with a single star polymer. 

\subsection{Star polymer}
As a first example of our formalism we consider a star consisting of a black
core out of which emerge  $f$ branches of $k$ white beads each. Evidently, in
this case the black system consists of a single bead, with the unique eigenvalue $\lambda_b=0$; then $P_{f,s}(\lambda)$,
Eq. (\ref{text_16}), does not appear. Furthermore, from Eq. (\ref{delta+}) one has $\Delta_+=0$, by inserting $N=1$ and $l=0$ into it. Apart from $\lambda_1=0$ there are two classes of non-trivial solutions: one class stems from $Q_{f,s}(\lambda)$ and in it the black bead moves (the modes are called
{\it symmetric} in the literature \cite{zimm}, see Refs. \cite{cai,gurtov_new} for the situation for dendrimers); in the other class, stemming from the $\Delta_-$ case, the black bead
is at rest (this leads to {\it antisymmetric }modes \cite{zimm}). From our formalism (and
as it is well-known) the symmetric modes are non-degenerate; there are $k$ of
them and they fulfill Eq. (\ref{Q=0}). The antisymmetric modes are, on the other hand,
$\Delta_-=(f-1)$ - fold degenerate; they lead each to the $k$ distinct eigenvalues
given by Eq. (\ref{text_18b}). For the star the reason for the degeneracy of the antisymmetric modes is
evident; it can be seen by pairwise combining in linearly independent manner, the $f$ branches of the star.

\subsection{The Cube polymer}
\begin{figure}[h]
\begin{center}
\epsfig{file=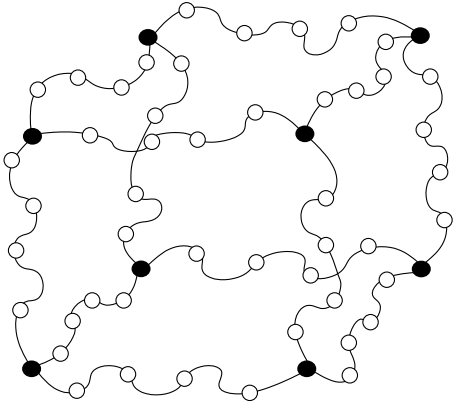,width=6cm}
\end{center}
\caption{The Cube polymer. As a general remark on its structure in solution, see caption of Fig.\ref{linked_stars}}
\label{cube_poly}
\end{figure}

Another very interesting example for the use of our formalism in a simple context is provided by a connected structure topologically identical to a cube. The basic network (black system) consists here of eight black beads which are the cores of eight stars with $f=3$ arms each and with $k$ white beads on each arm. The arms are then connected pairwise, through $l=12$ links, see Fig. (\ref{cube_poly}). This leads to 12 spacers of $s=2k$ white beads along the ``edges'' of the ``cube''.

The cube contains loops of spacers but no dangling bonds. Moreover, the structure is alternating, in the sense of Appendix C. Hence, a very interesting aspect of the cube is that it provides a very nice example of the special case treated in Appendix C: A rare, additional degeneracy related to $P_{f,s}(\lambda)$ and to $U_s=0$ occurs.

For the cube the eigenvalues of the black system are 6,4,2 and 0. Of these, $\lambda_b=0$ and $\lambda_b=6$ are non-degenerate, whereas $\lambda_b=4$ and $\lambda_b=2$ are each three-fold degenerate. Thus, our formalism leads, based on Eq. (\ref{text_16}), to $2k+1$ solutions to $P_{f,s}(\lambda)=6$ and to $3(2k+1)$ solutions to $P_{f,s}(\lambda)=4$ and $P_{f,s}(\lambda)=2$, each. We also find from Eq. (\ref{Q=0}) $k+1$ solutions corresponding to the equation $Q_{f,s}(\lambda)=0$. Furthermore, from Sec. 3.2 we compute, based on Eqs. (\ref{delta+}) and (\ref{delta-}), the multiplicities of the $k$ modes to Eqs. (\ref{36}) and (\ref{35}) to be $\Delta_+=5$ and $\Delta_-=4$. All in all, we thus obtain $7(2k+1)+(k+1)+9k=8(3k+1)$ modes. 

We remark now that for the cube one can construct a loop on every of the six faces, but that from these six loops, there are only five which are linearly independent. As far as each of these loops has an even number of chains, it can accommodate modes corresponding to Eq. (\ref{text_18b}) that are symmetric at mid-spacer. Thus, the total multiplicity of these modes is $\Delta_-+1=5$. As shown in Appendix C, the additional multiplicity of these $k$ modes is already accounted for by $k$ of the solutions to $P_{f,s}(\lambda)=2f=6$. For these $k$ modes, the factor $U_{2k}$ vanishes, see Appendix C, meaning that the black beads are at rest. We recall that this is a general feature of fully connected networks, in which all loops have an even number of chains, see Appendix B.

\subsection{Periodic networks}
As a third example we now apply our formalism to a topologically regular $d$-dimensional network, as it is treated in detail for $d=3$ in Ref. \cite{gurtov2}.

We start with a periodic, lattice-like arrangement of $N=N^{(1)}\times N^{(2)}\times\cdots\times N^{(d)}$ black beads and impose periodic boundary conditions. By this {\it all} the black nodes get to be $2d$-coordinated, $f=2d$. The topological regularity of the system of black beads allows readily to determine \cite{gurtov2} its eigenvalues $\lambda_b$. One finds:
\begin{equation}
\label{l_b}
\lambda_b=2d(1-\cos\Theta_b),
\end{equation}
where $\cos\Theta_b$, the so-called generating function is
\begin{equation}
\label{cos_tetha}
\cos\Theta_b=\frac{1}{d}\sum_{1=1}^{d}\cos\vartheta_i,
\end{equation}
with $\vartheta_i=S_i(2\pi)/N^{(i)}$ and $S_i=0,\ldots,N^{(i)}-1$. One may note that Eq. (\ref{l_b}) provides all the $N^{(1)}\times N^{(2)}\times\cdots\times N^{(d)}$ eigenvalues of the black system and that the {\it zero} eigenvalue, $\lambda_1=0$, corresponds to the choice $S_1=S_2=\ldots =S_d=0$.

We now replace all bonds of the black network by spacers with $2k$ white beads each. At this stage one has to solve Eq. (\ref{text_16}) to obtain the first class of modes of the whole (black and white) network. Now, as has been noted in Ref. \cite{gurtov2}, a considerable simplification of the problem occurs when the coefficient $\sigma$ happens to equal $f/2$, i.e. for $\sigma=f/2=d$. One may see the idea clearly by starting from Eq. (\ref{text_13}) for this parameter choice and recalling Eq. (\ref{2x}):
\begin{eqnarray}
\nonumber
P_{f,s}(\lambda) & = & f[1-(1-\frac{\lambda}{2})U_s+U_{s-1}] \\
\label{net_P}
& = & f(1-xU_s+U_{s-1}).
\end{eqnarray}
The condition Eq. (\ref{text_16}) is then, with Eq. (\ref{l_b}):
\begin{equation}
\label{51a}
\cos\Theta_b=xU_s-U_{s-1}.
\end{equation}
Now, using Eq. (\ref{text_9}), it follows
\begin{eqnarray}
\nonumber
\cos{\Theta_b} & = & \{\cos(\varphi)\sin[(s+1)\varphi]-\sin{(s\varphi)}\}/{\sin\varphi}= \\
\label{cos_theta_b}
& = & \cos[(s+1)\varphi],
\end{eqnarray}
since
\begin{eqnarray}
\nonumber
& & \sin{(s\varphi)}=\sin[(s+1)\varphi-\varphi]= \\
\label{52a}
&&\sin[(s+1)\varphi]\cos\varphi-\cos[(s+1)\varphi]\sin\varphi.
\end{eqnarray}
Equation (\ref{cos_theta_b}) is readily solved, by which we recover the solutions of Ref. \cite{gurtov2}:
\begin{equation}
\label{the_solutions}
\varphi_m=\frac{2m\pi}{s+1}\pm\frac{\Theta_b}{s+1},\hspace{1cm}\textrm{with}~~m=0,1,2,\ldots,k,
\end{equation}
where one has to keep only one solution for $m=0$. Excluding the case $\lambda_b=0$, Eqs. (\ref{cos_tetha}) and (\ref{cos_theta_b}) lead to $N_P=(N-1)(s+1)$ eigenvalues. 

We now turn to the degenerate eigenvalues corresponding to Eqs. (\ref{text_18b}) and (\ref{loops_b}). Using Eqs. (\ref{delta-}) and (\ref{delta-}) for $\Delta_+$ and $\Delta_-$ and recalling that we have here a highly connected periodic network, without dangling bonds, whose number of links is $l=dN$, we find as in Ref. \cite{gurtov2} $\Delta_-=(d-1)N$ and $\Delta_+=\Delta_-+1$.

It remains to study Eq. (\ref{Q=0}). Now for the choice $\sigma=f/2$ one again has with Eq. (\ref{Q_simplified}) the simplification
\begin{equation}
\label{Q_simplified_net}
Q_{f,s}(\lambda)=d\lambda(U_k+U_{k-1}),
\end{equation}
which leads to an additional, accidental increase of the number of eigenvalues corresponding to Eq. (\ref{loops_b}).

Finally we note that our approach is not limited to periodic boundary conditions, but can handle as well finite regular structures with dangling half chains.

\section{Dendrimers with spacers}
\begin{figure}
\begin{center}
\epsfig{file=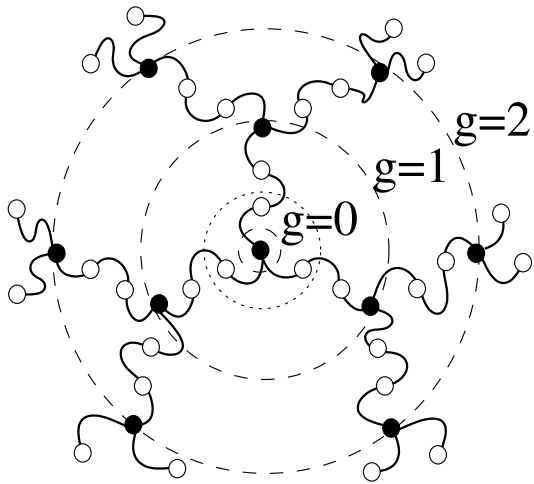,width=7cm}
\end{center}
\caption{A second generation dendrimer with spacers. Here, $g=2, f=3$ and each spacer chain contains two beads ($s=2, k=1$). The dash-dotted line indicates one of the stars from which the structure is built. As a general remark on the dendrimer's structure in solution, see caption of Fig. \ref{linked_stars}.}
\label{dendri_spacers}
\end{figure}

As a specific example of branched polymers, we treat here star burst dendrimers, which are built from $f$-functional stars, linked in such a way as to describe a regular Cayley tree, see Figure \ref{dendri_spacers}. One may remark that our renormalization procedure reduces the dendrimer with spacers and dangling bonds of Fig. \ref{dendri_spacers} to a simple Cayley tree. Now, let us recall the properties of the eigenvalue spectrum of such a simple dendrimer \cite{grassley,cai,wedges,gurtov_new,ferla,markelov_new}. As has also been noted in Ref. \cite{wedges}, this problem can be partly mapped to a rescaled one-dimensional Gaussian chain. Thus, it allows for a formulation in terms of Chebyshev polynomials \cite{kloczkowski}.

We start with the system of black beads, i.e. with the dendrimer at the generation $g$. The total number of black beads is then:
\begin{equation}
\label{N_total}
N=1+f\frac{(f-1)^g-1}{f-2}.
\end{equation}
As discussed in Refs. \cite{cai,wedges,gurtov_new,grassley} the eigenvalue spectrum corresponds to two different types of modes: those in which the core is mobile and those in which the core does not move.

Equations for the 
eigenvalues $\lambda_b$ of the classical dendrimer of generation $g$ and general functionality $f$ have been derived in Refs. \cite{cai,gurtov_new}. In terms of the Chebyshev polynomials of the second kind these eigenvalues can be obtained from the roots of the following equations, see Appendix D:
\begin{equation}
\label{mob_core}
U_{g}(x) +  \frac{1}{\sqrt{f-1}}U_{g-1}(x)=0 
\end{equation}
and
\begin{equation}
\label{immob_core}
U_{m}(x)  -  \sqrt{f-1}U_{m-1}(x)=0 ,~~\textrm{with}~m=1,\ldots,g.
\end{equation}
The eigenvalues themselves are then given by
\begin{equation}
\label{argument}
\lambda_b=f-2x\sqrt{f-1}.
\end{equation}
We remark that for $g+1\leq\sqrt{f-1}$ and for $m<(m-1)\sqrt{f-1}$ these solutions
include the so-called exponential modes with $|x|>1$, for which Eq. (\ref{text_9}) may
be written in terms of hyperbolic functions.  
The $g$ solutions to Eq. (\ref{mob_core}) are non-degenerate, while the degeneracy of the eigenvalues to Eq. (\ref{immob_core}) is given by 
\begin{eqnarray}
\label{degeneracy}
\Delta_{m}=\left\{ \begin{array}{ll}
(f-1) & \textrm{ if $m=g$} \\
f(f-2)(f-1)^{g-(m+1)} & \textrm{otherwise}.
\end{array}\right. 
\end{eqnarray}
 This defines the full eigenvalue spectrum of the dendrimer consisting of the black beads only.

Now, we calculate the eigenvalues of the full system (the system including spacers and dangling chains) by solving Eqs. (\ref{text_16}), (\ref{35}) and (\ref{Q=0}). Each eigenvalue $\lambda_b\ne 0$ of the black system gives rise to a set of $s+1$ eigenvalues of the full system which inherit its degeneracy. The degeneracy of the solutions to Eq. (\ref{35}) is $\Delta_-$, see Eq. (\ref{delta-}), while the solutions to Eq. (\ref{Q=0}) are non-degenerate.
\begin{figure}[h]
\begin{center}
\epsfig{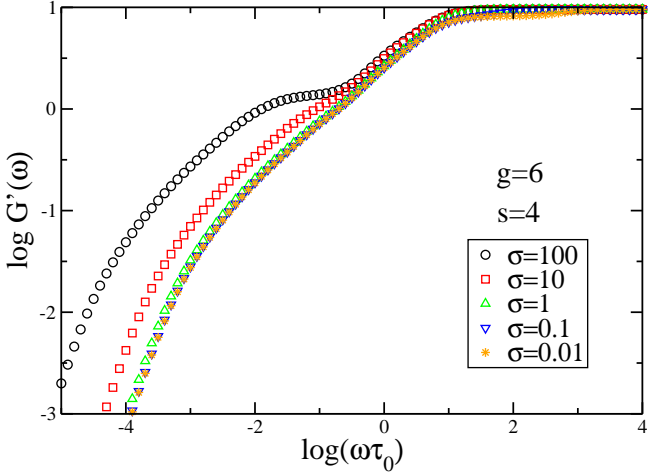}
\end{center}
\caption{Storage modulus $G'(\omega)$ for a dendrimer of generation $g=6$, functionality $f=3$, with spacer chains of length $s=4$ in between the branching points and dangling bonds of length $k=2$. Here, $\sigma$ varies from $0.01$ to $100$.}
\label{SDM_different_gama}
\end{figure}

This two-step approach  (of first finding the eigenvalue spectrum of the reduced black system, and only then calculating the eigenvalue spectrum of the full system, by solving Eqs. (\ref{text_16}), (\ref{35}) and (\ref{Q=0})), considerably reduces the complexity of the problem. We should note that the maximal degrees of the polynomial equations involved here are $g$ and $s+1$. As shown below, this allows us to numerically treat dendrimers with spacers having a total of more than $14000$ beads. This demonstrates one of the main advantages of our present work, when comparing it to a one-step approach \cite{kloczkowski}; such an approach leads here to a polynomial equation of much higher degree, namely of degree $(s+1)(g+1)$. We also note that in Ref. \cite{kloczkowski} the peripheral beads of the dendritic system were assumed to be fixed in space. In retrospect, this assumption, besides of being rather arbitrary, appears to be counterproductive.
\begin{figure}[h]
\begin{center}
\epsfig{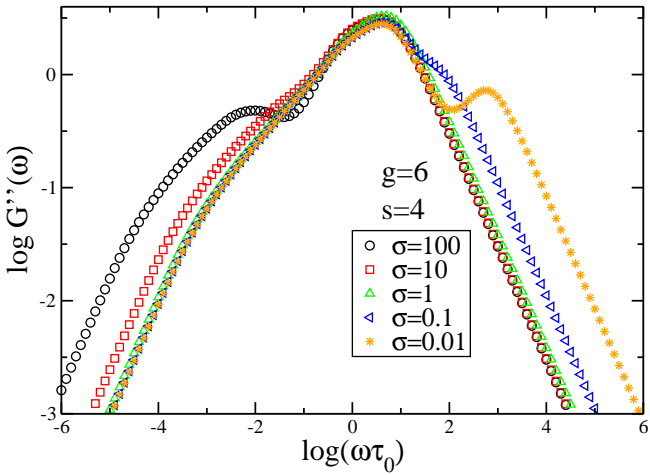}
\end{center}
\caption{Loss modulus $G''(\omega)$ for a dendrimer of generation $g=6$, functionality $f=3$, with spacer chains of length $s=4$ in between the branching points and dangling bonds of length $k=2$. Here, $\sigma$ varies from $0.01$ to $100$.}
\label{LDM_different_gama}
\end{figure}

Let us now demonstrate numerically the effects of varying the parameters $s=2k$ and $\sigma$, which determine the length of the spacers and of the dangling chains, and the ratio of the friction coefficients, respectively. We choose here a $g=6$ generation dendrimer with functionality $f=3$, and show its storage $G'(\omega)$ and loss $G''(\omega)$ moduli, calculated according to Eqs. (\ref{gprim}) and (\ref{gdprim}).

In Figs. \ref{SDM_different_gama} and  \ref{LDM_different_gama} we display $G'(\omega)$ and $G''(\omega)$ as functions of the reduced frequency $\bar\omega$ for dendrimers with chain length parameter $s=2k=4$ while varying $\sigma$ from $0.01$ to $100$. The mobility of the white beads (i.e. $\zeta_0$) is kept constant. The curve corresponding to $\sigma=1$ describes the case when all the beads in the system have the same mobility. As it was shown \cite{sgb1,sgb2}, differences in the  mobilities of the different kinds of beads lead to the appearance of plateau-type behaviors in $G'(\omega)$ and to additional peaks in $G''(\omega)$. For $\sigma>1$ the spacers and dangling beads are the ``lighter'' beads, being more mobile than the branching units (black beads). Their number being six times larger, their relaxation dominates the spectrum. In particular, for $\sigma=100$, the loss modulus $G''(\omega)$ displays a major peak at intermediate frequencies, corresponding to the relaxation of the white beads and a second minor peak at lower frequencies, corresponding to that of the black beads. For $\sigma<1$ the situation is reversed: The black beads are more mobile and the minor peak corresponding to their relaxation is shifted to higher frequencies, while the major peak induced by the white beads keeps its position. Corresponding effects are seen also in the behavior of the storage moduli $G'(\omega)$, which for $\sigma\ne 1$ display a plateau resulting from a gap in the relaxation. The plateau is formed between the domains of high and of low frequency relaxation. For $\sigma=100$ the white beads dominate the high frequency relaxation, leading to a lower level of the plateau than for $\sigma=0.01$, where the high frequency relaxation of the black beads is only weakly evident.

\begin{figure}[h]
\begin{center}
\epsfig{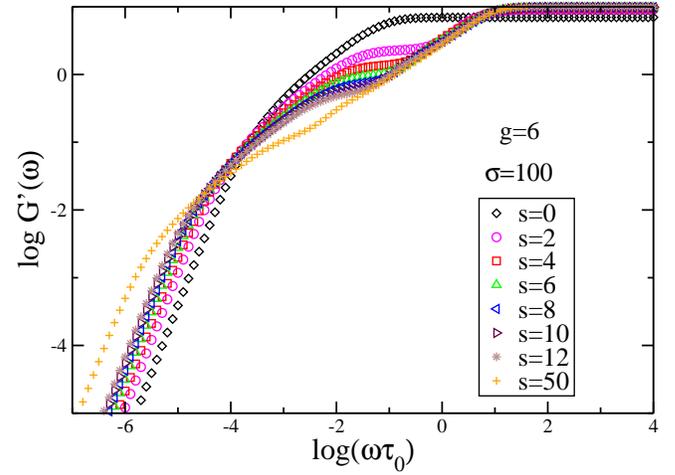}
\end{center}
\caption{The storage modulus for a dendrimer of generation $g=6$, when the number of spacers between the branching points grows from $s=2$ to $s=50$. In all the situations $\sigma=100$.}
\label{SDM_different_spac}
\end{figure}
\begin{figure}[h]
\begin{center}
\epsfig{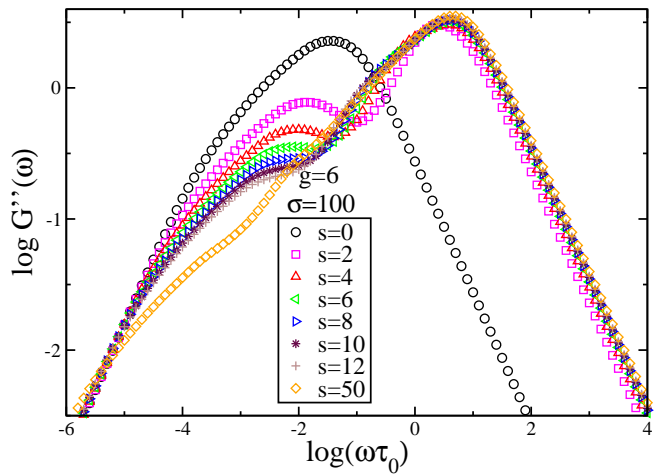}
\end{center}
\caption{The loss modulus for a dendrimer of generation $g=6$, when the number of spacers between the branching points grows from $s=2$ to $s=50$. In all the situations $\sigma=100$.}
\label{LDM_different_spac}
\end{figure}

To demonstrate the influence of increasing the length of the spacers and of the dangling chains,  we plot in Figs. \ref{SDM_different_spac} and \ref{LDM_different_spac} the corresponding results for the moduli $G'(\omega)$ and $G''(\omega)$. Again we consider a $g=6$ generation dendrimer with functionality $f=3$. Fixing the parameter $\sigma=100$ (higher mobility of the white beads) we vary the length of the spacer and of the dangling chains, by increasing $s=2k$ from $s=0$ to $s=50$. For $s=50$, the full system consists of $N_{full}=14440$ beads, while the reduced (black) system is a dendrimer containing $N=190$ beads. The classical dendrimer ($s=0$) shows a typical, non-scaling behavior of the relaxation part of the spectrum, both for $G'(\omega)$ and for $G''(\omega)$. Increasing $s$ the relaxation moduli display a mixture of dendrimer and chain like behavior. For short chains we observe in $G''(\omega)$ a second minor peak at lower frequencies, which, for longer chain lengths becomes less and less pronounced, remaining eventually only visible as a shoulder. For long chains both $G'(\omega)$ and $G''(\omega)$ are dominated by the chain like relaxation behavior of the white beads. 

\section{Conclusions}
The dynamics of many branched polymer systems can be understood in terms of the dynamics of end-linked star polymers. Here, we presented a general method to analytically treat this problem, while also including the important effects brought along by changes in the length and in the mobility of the arms of the stars. We developed a mathematically exact procedure which allows to determine the full eigenvalue spectrum of such networks, by exactly mapping the eigenmode spectrum of a network formed by end-linking star polymers to that of a reduced, core system. Namely, for any such network (under the weak restriction that adjacent stars are linked by one bond only) one obtains the core system by replacing each star polymer by its core. Given the eigenvalue spectrum of this core system we obtain the full spectrum of eigenvalues by solving a set of polynomial equations. For a network of star polymers with $f$ arms of $k$ beads each, every non-zero eigenvalue of the core system gives rise to $2k+1$ non-trivial eigenvalues of the full network. These constitute the first class of eigenvalues; the second class is related to the zero eigenvalue of the core system and is organized in three groups: The first two groups correspond to the symmetric and antisymmetric modes of the arms of a single star polymer, while the third group corresponds to the modes of spacer chains with fixed ends. For all these modes we determined their eigenvalue multiplicities; these depends only on the topology of the core system, i.e. on the total number of stars and on the number of independent loops in the core system, as worked out in Appendix B. 

We demonstrated the full power of our approach by recovering readily the full spectra of model structures such as star polymers and regular $d$-dimensional networks; we reproduced former results for $d=3$, and generalized these to arbitrary $d$. Furthermore, we exemplified hidden symmetries of the problem using the ``cube'' polymer as fundamental example.  Additionally, we performed extensive calculations on dendritic structures containing spacers and dangling chains. Our two-step approach allowed us to readily treat systems containing more than $14000$ beads; in fact, numerically we are limited only by our possibility of determining the spectra of the core system; this means for arbitrary systems, of some $5000$ black beads, and (again) practically unlimited for hierarchical (fractal, dendritic, regular) core networks. We note that a previous approach to dendrimers with spacers \cite{kloczkowski}, restricted to systems with fixed peripheral junctions, did not succeed in decoupling the problem and obtained only approximate spectra (sometimes, even negative, unphysical eigenvalues).

Considering the relaxation moduli of copolymers, we find strong effects when the mobilities of the beads involved differ. Also, increasing the length of the arms of the stars leads to smooth transitions between the relaxation behavior due to the core and due to the linear chains.

Our method is very general, in that the topology of the core and the details of the stars' geometries are (in a large measure) arbitrary. It provides a scheme which considerably reduces the calculational efforts, by splitting the task into two steps. Furthermore, if any symmetry is present in the core system, additional reduction steps are possible. As an outlook, we note that the method presented here allows to generalize recent analytic results for the dynamics of randomly branched polymers \cite{jasch,ferber,jfb} to systems built from them by the insertion of spacers and of dangling bonds, and in which the mobility of the beads may vary.

\begin{acknowledgments}
We thank Andrey A. Gurtovenko for discussions and for a critical reading of the manuscript. The help of the Deutsche Forschungsgemeinschaft and of the Fonds der Chemischen Industrie is gratefully acknowledged.
\end{acknowledgments}

\appendix
\section{}

In this Appendix we derive several relations connecting the Chebyshev polynomials of the second kind, $U_n(x)$, Eqs. (\ref{text_3}) and (\ref{text_9}) of the main text. We start from the relation:
\begin{eqnarray}
\nonumber
&&\sin{(2n\varphi)}\sin \varphi = 2\sin{(n\varphi)}\cos{(n\varphi)}\sin \varphi= \\
\nonumber
& &= \sin{(n\varphi)}[~\sin{(n\varphi+\varphi)}-\sin{(n\varphi-\varphi)}~] \\
\label{sinus}
& &= \sin{(n\varphi)}\{\sin{[(n+1)\varphi]}-\sin{[(n-1)\varphi]}\}
\end{eqnarray}
Using Eq. (\ref{text_9}), it follows by division with $\sin^2\varphi$ that
\begin{equation}
\label{U_2n-1}
U_{2n-1}=U_{n-1}(U_{n}-U_{n-2}).
\end{equation}
Moreover from:
\begin{eqnarray}
\nonumber
&&\sin{[(n+1)\varphi]}\sin{[(n-1)\varphi]}= \\
\nonumber
&& =[\cos \varphi\sin{(n\varphi)}+\cos(n\varphi)\sin \varphi]\times \\
\nonumber
& &\times [\cos \varphi\sin(n\varphi) - \cos(n\varphi)\sin \varphi]= \\
\nonumber
& &= [\cos^2{\varphi}\sin^2(n\varphi)-\cos^2(n\varphi)\sin^2{\varphi}] \\
\nonumber
& &= \cos^2{\varphi}\sin^2(n\varphi)-[1-\sin^2(n\varphi)](1-\cos^2{\varphi}) \\
\nonumber
& &= \cos^2{\varphi}\sin^2(n\varphi)-\sin^2{\varphi}+\sin^2(n\varphi)-\cos^2{\varphi}\sin^2(n\varphi) \\
\label{sin^2(nvarphi)}
& &= \sin^2(n\varphi)-\sin^2{\varphi},
\end{eqnarray}
it follows by division with $\sin^2 \varphi$ that 
\begin{equation}
\label{sin_square}
U_nU_{n-2}=U_{n-1}^2-1,
\end{equation}
and hence 
\begin{equation}
\label{U^2_n-1}
U_{n-1}^2-U_nU_{n-2}=1.
\end{equation}
Thus, from Eqs. (\ref{U_2n-1}) and (\ref{U^2_n-1})
\begin{eqnarray}
\nonumber
&&(U_n+U_{n-1})(U_{n-1}-U_{n-2})=U_nU_{n-1}+U_{n-1}^2-\\
\label{old22}
&&-U_nU_{n-2}-U_{n-1}U_{n-2}=U_{2n-1}+1.
\end{eqnarray}
Furthermore, one also has
\begin{equation}
\label{old21}
(U_n+U_{n-1})(U_n-U_{n-1})  =  U_n^2-U_{n-1}^2=U_{2n} ,
\end{equation}
as we proceed to show based on the relation $$\sin^2\varphi_1-\sin^2\varphi_2=\sin{(\varphi_1+\varphi_2)}\sin{(\varphi_1-\varphi_2)},$$ Eq. (4.3.20) of Ref.\cite{Abramovitz}. Thus
\begin{equation}
\label{trigonometric}
\sin^2[(n+1)\varphi]-\sin^2(n\varphi)=\sin[(2n+1)\varphi]\sin\varphi.
\end{equation}
Dividing by $\sin^2\varphi$ and applying Eq. (\ref{text_9}) proves now Eq. (\ref{old21}).

\section{}

In this Appendix we derive the multiplicities, Eqs. (\ref{delta+}) and (\ref{delta-}) of the modes to Eqs. (\ref{35}) and (\ref{36}), by an explicit construction. Let $\hat\Phi$ and $\hat\Psi$ be modes on a spacer with $2k$ white beads with fixed end black beads, to given eigenvalues, Eqs. (\ref{35}) and (\ref{36}), where $\hat\Phi$ is symmetric and $\hat\Psi$ is antisymmetric at mid-spacer:
\begin{equation}
\label{star}
\hat\Phi_k=\hat\Phi_{k+1}, \hspace{0.6cm} \mbox{and} \hspace{0.6cm} \hat\Psi_k=-\hat\Psi_{k+1}. 
\end{equation}
\begin{figure}[ht]
\begin{center}
\epsfig{file=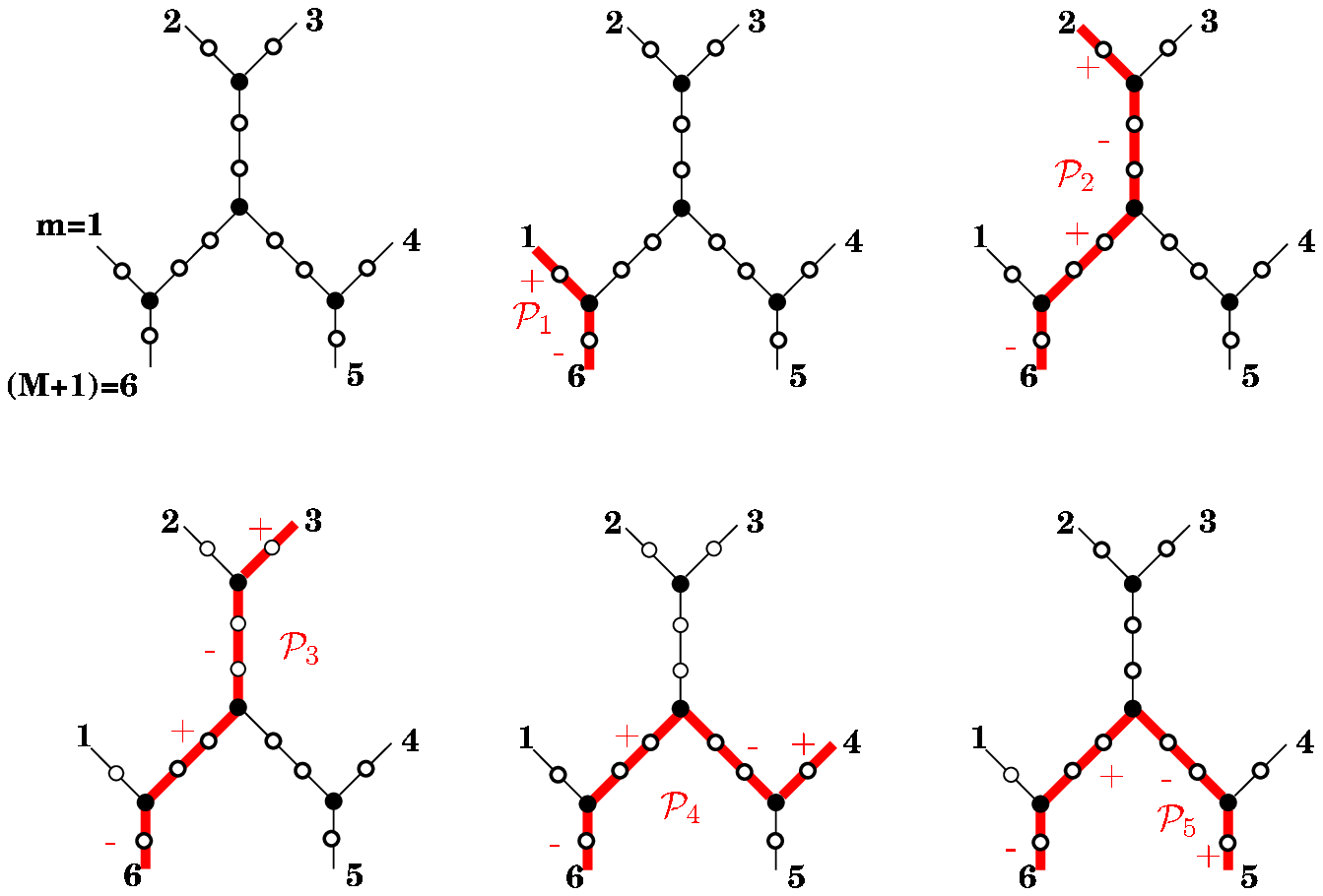,width=0.99\columnwidth}
\end{center}
\caption{Directed paths $\mathcal{P}_1,\ldots ,\mathcal{P}_5$ between the dangling chain $(M+1)=6$ and $m$ ($m=1,\ldots ,5$) and the corresponding modes $\Phi^{(m)}\Big\vert_a=\pm\hat\Phi$ for a structure obtained by end-linking $N=4$ polymer stars. Indicated by $+/-$ signs are the amplitudes of the modes $\Phi^{(m)}$.}
\label{paths}
\end{figure}
Consider now one full structure (with white and black beads), formed from end-linking $N$ stars by $l$ links, $l\ge l_1=N-1$. If $l_1=l$ then the structure is topologically a tree and the total number $M+1$ of dangling chains is $M+1=(f-2)N+2$. For $l>l_1$ the network contains loops and we set $L=l-l_1$, so that $l=L+N-1$. By removing now from the structure $L$ judiciously chosen links out of the $l$ links, the ensuing graph turns out to be a connected tree. We number then the new dangling chains (created by removing the $L$ links) by $(1,2),(3,4),\ldots ,(2L-1,2L)$ and the other dangling chains of the tree (these exist for $2L\le M-1$) by $2L+1,\ldots ,M+1$. 

We now assume $2L\le M$ (then $M+1$ is a dangling chain of the original structure) and specify $M$ paths $\mathcal{P}_m$, where $\mathcal{P}_m$ is the directed path between the dangling chains  $M+1$ and $m$ (with $m=1,\ldots ,M$). Since our structure is a tree, each directed path is unique. For each path $\mathcal{P}_m$ we define a mode $\Phi^{(m)}$ by its values when restricted to any (half-) spacer $a\in \mathcal{P}_m$:
\begin{equation}
\label{phim}
\Phi^{(m)}\Big\vert_a=\pm\hat\Phi,
\end{equation}
where the sign oscillates between adjacent spacers along the path (see Figure \ref{paths}). Here we use an implicit numbering (in the direction of the path $\mathcal{P}_m$) of the beads along each spacer, with obvious restrictions for the dangling chains. We fix the $+$ sign for the dangling chain $m$. Outside the path $\mathcal{P}_m$, the mode $\Phi^{(m)}$ vanishes. Each mode $\Phi^{(m)}$ is formed by symmetric pieces along the spacer and fulfills Eq. (\ref{text_10}) at the black beads. The modes $\Phi^{(m)}$ are obviously linearly independent, since each path $\mathcal{P}_m$ contains at least one segment that is not contained in any other path. In the case $\Delta_+=0$ we have thus constructed $M=(f-2)N+1$ different degenerate modes corresponding to a given eigenvalue of Eq. (\ref{text_18b}), by which we have proven the correctness of Eqs. (\ref{delta+}) and (\ref{delta-}) in the case of a tree structure. 

\begin{figure}[ht]
\begin{center}
\epsfig{file=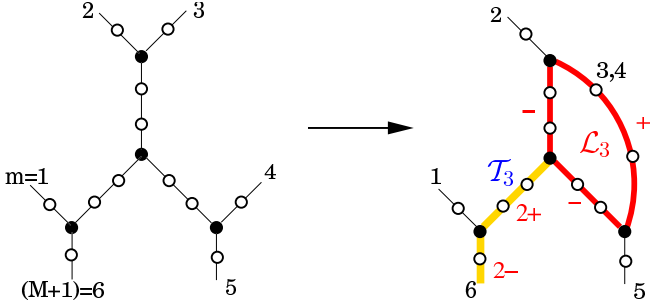,width=0.99\columnwidth}
\end{center}
\caption{The loop $\mathcal{L}_3$ and tail $\mathcal{T}_3$ obtained when introducing one additional link between the dangling chains $m=3$ and $m=4$. We also indicate the amplitudes of the mode $\Phi^{(3,4)}$ by $+/-$ signs.}
\label{loops_fig}
\end{figure}

In the following we start to reconnect the dangling chains $(1,2),(3,4),$ etc. and follow the evolution of the system of eigenmodes under this operation. For this we define for each path $\mathcal{P}_m$ on the tree a mode $\Psi^{(m)}$ by its restrictions
\begin{equation}
\label{psim}
\Psi^{(m)}\Big\vert_a=\hat\Psi, \hspace{0.6cm}\mbox{with}\hspace{0.4cm} a\in \mathcal{P}_m,
\end{equation} 
while we let it vanish outside $\mathcal{P}_m$. Note, however, that, due to its eigenvalue, $\Psi^{(m)}$ cannot involve the motion of dangling chains, a fact which we ignore for the moment. Furthermore, we construct the modes:
\begin{equation}
\label{2stars_a}
\Phi^{(m,m+1)}  =  \Phi^{(m)}+\Phi^{(m+1)},
\end{equation}
and
\begin{equation}
\label{2stars}
\Psi^{(m,m+1)}  =  \Psi^{(m)}-\Psi^{(m+1)}, 
\end{equation}
with $ m=1,3,\ldots,2L-1$
The union of paths $\mathcal{P}_m\cup \mathcal{P}_{m+1}$, consists of a loop $\mathcal{L}_m$ and a ``tail'' $\mathcal{T}_m$ between this loop and the dangling chain $M+1$ given by the intersection $\mathcal{P}_m\cap \mathcal{P}_{m+1}$ (see Figure \ref{loops_fig}). We will call $\mathcal{L}_m$ an even or odd loop, if it contains an even or odd number of spacers.

By construction, the modes $\Phi^{(m,m+1)}$ and $\Psi^{(m,m+1)}$ fulfill the boundary condition (\ref{star}) at the center of the spacer and Eq. (\ref{text_10}) at the branching point of paths $\mathcal{P}_m$ and $\mathcal{P}_{m+1}$. Mode  $\Psi^{(m,m+1)}$ is non-zero only on the loop $\mathcal{L}_m$, whereas mode $\Phi^{(m,m+1)}$ is also non-zero on the tail $\mathcal{T}_m$ if $\mathcal{L}_m$ is odd. Note that each linkage (reconnection) reduces the number of $\Phi$-type eigenmodes by unity and increases the number of $\Psi$-type eigenmodes by unity. By insertion of $L$ links one obtains $L$ modes belonging to Eq. (\ref{2stars_a}), $L$ modes belonging to Eq. (\ref{2stars}), but has now only $M-2L$ modes $\Phi^{(m)}$ on the paths $\mathcal{P}_m$ ($m>2L$) connecting dangling chains. We obtain thus:
\begin{equation}
\label{delta-+}
\Delta_+ = L 
\end{equation}
and
\begin{equation}
\label{delta+-}
\Delta_-  = (M-2L)+L=M-L.
\end{equation}
Given that $M=(f-2)N+1$ and $l=L+N-1$, these expressions are equivalent to Eqs. (\ref{delta+}) and (\ref{delta-}). 

We turn now to the case which we excluded up-to-now, namely to $2L=M+1$. Then the original structure has no dangling chains. In this case we perform first the steps above, and reconnect $(1,2),(3,4)\ldots$ up to $(M-2,M-1)$. By this only the path $\mathcal{P}_{M}$ between the dangling chains $M$ and $M+1$ remains to be closed. Connecting these two dangling chains, the path $\mathcal{P}_{M}$ becomes a loop, which we denote by $\mathcal{L}_M$.

\begin{figure}[ht]
\begin{center}
\epsfig{file=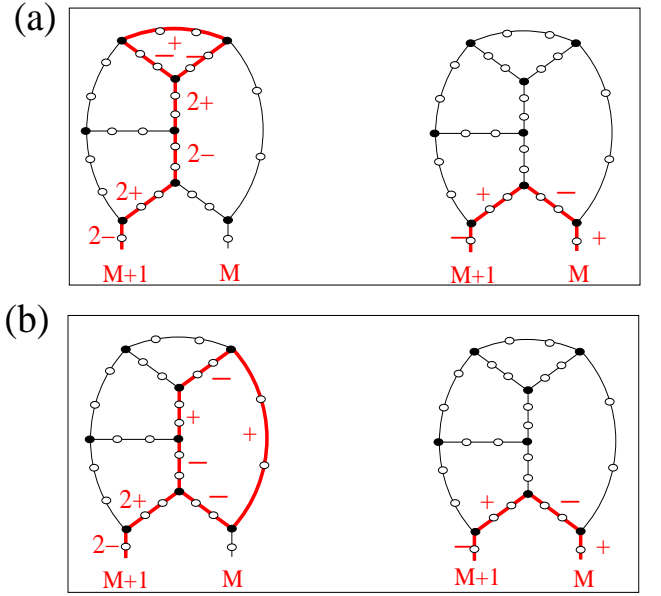,width=0.99\columnwidth}
\end{center}
\caption{Two examples of situations encountered while constructed the modes $\tilde\Phi^{(j,j+1)}$ when closing the last link $(M,M+1)$.}
\label{last_loop}
\end{figure}
First, we assume that some of the loops $\mathcal{L}_j$ for modes
corresponding to Eq. (\ref{2stars}) have an odd number of spacers
(Figure \ref{last_loop}). Let the paths be ordered such that
$\mathcal{L}_M$ is odd. The above considerations remain valid for all
modes $\Psi^{(m,m+1)}$ as well as for all modes $\Phi^{(m,m+1)}$ with
an even $\mathcal{L}_m$. However, upon closing the last link
transforming the path ${\cal P}_M$ to a loop ${\cal L}_M$, all modes
$\Phi^{(j,j+1)}$ that include a nonvanishing tail part ${\cal T}_j$,
i.e. those for which $\mathcal{L}_j$ is odd, have to be redefined.  The
last link consists of the two dangling half chains $M$ and $M+1$. The
amplitudes of a given mode $\Phi^{(j,j+1)}$ on the chains along the
tail ${\cal T}_j$ are $\pm 2$: hence on the dangling chain $M+1$ the mode
takes the value $2\epsilon_j$, with either $\epsilon_j=+1$ or $\epsilon_j=-1$. On the
dangling chain $M$ the amplitude of $\Phi^{(j,j+1)}$ vanishes; closing the loop ${\cal L}_M$ we are faced with an amplitude mismatch of
$2\epsilon_j$ between the dangling chains $M$ and $M+1$, fact which we have to mend.  Now, since here $\mathcal{L}_M$
is an odd loop, the mode  $\Phi^{(M)}$ defined on $\mathcal{P}_M$
has amplitude $+1$ on the dangling chain $M$ and amplitude $-1$
on the dangling chain $M+1$. Thus, closing the loop $\mathcal{L}_M$,
the mode $\Phi^{(M)}$ will exhibit an amplitude mismatch of $\delta=2$
between the dangling chains $M$ and $M+1$, while being well defined
everywhere else. This fact allows us to construct for each odd loop
$\mathcal{L}_j$ a well defined $\Phi$-type mode (which lives on 
the union of $\mathcal{L}_j$, $\mathcal{T}_j$ and $\mathcal{L}_M$) by
\begin{equation}
\label{system}
\tilde\Phi^{(j,j+1)}=\Phi^{(j,j+1)}+\epsilon_j\Phi^{(M)},
\end{equation}
where $2\epsilon_j$ is the amplitude of mode $\Phi^{(j,j+1)}$ on the
dangling chain $M+1$.
For all $j< M$ with an odd loop $\mathcal{L}_j$ these modes fulfill
Eq. (\ref{text_10}) and all other boundary conditions. On the other
hand, $\Psi^{(M)}$ becomes a valid mode on the loop
$\mathcal{L}_M$. Thus, as before, the closure of a loop leads here to
the disappearance of one $\Phi$-type mode $\Phi^{(M)}$ (counted by $\Delta_-$)
and to the creation of a $\Psi$-type mode (counted by $\Delta_+$). Hence
Eqs. (\ref{delta-+}) and (\ref{delta+-}), and thus Eqs. (\ref{delta+})
and (\ref{delta-}) hold.

Finally, we concentrate on the special situation when all loops $\mathcal{L}_m$ are even. Then, when closing the last link, all modes (\ref{2stars}) remain valid, since no nonvanishing tail parts have to be considered. Formally, the closure of the loop allows now to define {\it two} new modes, see Eqs. (\ref{2stars_a}) and (\ref{2stars}), one of $\Phi$- and one of $\Psi$-type, which replace the previous $\Phi^{(M)}$-mode. However, as shown in Appendix C, the additional $\Phi$-type  mode which would increase $\Delta_-$ by unity and would lead to $k$ distinct modes of the whole, black and white system, is only apparently new. All these seemingly new $k$ modes are in fact already included in the solutions of Eq. (\ref{text_16}), as we prove in Appendix C.

\section{}

In this Appendix we study whether for certain $\lambda_b\ne 0$ the relation $P_{f,s}(\lambda)=\lambda_b$ may have solutions related to $U_s=0$. Here we show that this can occur only if $\lambda_b=2f$. Now, from Eq. (\ref{text_9}), $U_s(\cos\varphi_r)=0$ means that
\begin{equation}
\label{U_s=0}
\varphi_r  =\frac{r\pi}{2k+1} \hspace{1.5cm} \textrm{with}~~ r=1,2,\ldots,2k,
\end{equation}
from which follows that
\begin{eqnarray}
\nonumber
U_{s-1}(\cos{\varphi_r}) & = & \frac{\sin{[(2k+1)\varphi_r-\varphi_m]}}{\sin{\varphi_r}}  \\
\label{U_s-1}
& = & \frac{\sin{(r\pi-\varphi_r)}}{\sin{\varphi_r}}=\left\{ \begin{array}{ll}
+1 & \textrm{for {\it r} odd} \\
-1 & \textrm{for {\it r} even}.
\end{array}\right.
\end{eqnarray}
For  $U_s=0$, $U_{s-1}$ takes one of these two possible values. Making now use of Eqs. (\ref{text_13}) and (\ref{text_16}) we see that:
\begin{equation}
\label{lambda_hom_appendix}
\lambda_b=f+fU_{s-1}=\left\{ \begin{array}{ll}
2f & \textrm{for}~ r=2m-1,~\textrm{with}~ m=1,\ldots,k \\
0 & \textrm{for}~ r=2m,~\textrm{with}~ m=1,\ldots,k.
\end{array}\right.
\end{equation}
Hence, the only possibility is to take $\lambda_b=2f$, given that we are looking for solutions to $P_{f,s}(\lambda)=\lambda_b$ with $\lambda_b\ne 0$.

To show now that indeed $\lambda_b=2f$ is a possible eigenvalue in the system consisting of the black beads only, we analyze the existence and the properties of an eigenmode to it. We start with Eq. (\ref{text_12}) for the system of black beads, where we consider a general site $i$ and its $f_i$ neighbors
\begin{eqnarray}
\label{A19a}
(f_i-2f)\Phi_i & = & \sum_{\alpha=1}^{f_i}\Phi_{i_{\alpha}} \\
\nonumber
\hspace{-6cm}\textrm{i.e.}~~~~~~~~~~~~~~~~~~~~~~~~~~~~~~~~~~~~~~~~~~~~~~~~~~~~~~~~~~~~~ & & \\
\label{lambda_equality}
2f\Phi_i & = & \sum_{\alpha=1}^{f_i}(\Phi_i-\Phi_{i_{\alpha}}).
\end{eqnarray}
We now pick the site $j$ such that on it $\vert\Phi_j\vert$ is maximal, $\vert\Phi_j\vert=max_i\vert\Phi_i\vert$. Furthermore we choose the amplitudes such that $\Phi_j$ is positive. Then from Eq. (\ref{lambda_equality})
\begin{equation}
\label{lambda_ineq}
2f\Phi_j=\sum_{\alpha=1}^{f_i}(\Phi_j-\Phi_{j_{\alpha}})\le 2f_j\cdot \Phi_{j}.
\end{equation}
This means that, on one side $2f\le 2f_j$, whereas, by construction, $f_j\le f$. It follows that we must have $f_j=f$.

Moreover, inserting $f_j=f$ into Eq. (\ref{lambda_ineq}) it also follows that we must have $\vert\Phi_j-\Phi_{j_{\alpha}}\vert=2\vert\Phi_j\vert$ for all $\alpha$, which simply implies $\Phi_{j_{\alpha}}=-\Phi_j$ for all the neighbors of $j$. Obviously, this argument can be repeated for all the $f$ neighbors of $j$, etc. Hence, the only mode compatible with $\lambda_b=2f$ is an alternating mode of two sublattices moving against each other. Such a mode is only possible when: first, all black beads have the functionality $f$ {\it on the black lattice}, and secondly, {\it all} loops in the structure are even. One should note that the first condition means that no dangling chains are allowed. 

We now turn to the motion of the full lattice (white and black beads); we recover it by replacing each bond of the black lattice with a spacer of $2k$ white beads. From Eq. (\ref{lambda_hom_appendix}) the condition $\lambda_b=2f$ implies that the $r$ must be odd, $r=2m-1$, and that there are exactly $k$ solutions satisfying this condition. Furthermore, based on Eq. (\ref{new2}) it follows that all these solutions correspond to eigenmodes which are even. Symmetric (even) eigenmodes are, however, compatible with having $\Phi_0=\Phi_{2s}$; on the other hand, from $\lambda=2f$, we have established that we must have $\Phi_0=-\Phi_{2s}$. It follows that $\Phi_0=\Phi_{2s}=0$, i.e. that in the $\lambda=2f$ mode the insertion of spacers that move symmetrically along their bonds leads to the immobilization of the black beads. We hasten to note that these eigenmodes (involving white and black beads) have the same characteristics as the class of eigenmodes discussed in Appendix B.

Given the relative complexity of the above analysis we have verified it by considering a series of different, alternating black lattices, and by numerically computing the eigenmodes and eigenvalues of the corresponding black and white systems, obtaining perfect agreement in all cases.

\section{}
Here, we show how the results for the eigenmode spectrum of generalized
dendrimers derived in Refs. \cite{cai} and \cite{gurtov_new} can be simplified by rewriting them in terms of Chebyshev polynomials as displayed in  
Eqs. (\ref{mob_core}) and (\ref{immob_core}) of the main text.
 Following Ref. \cite{gurtov_new}, the non-trivial
eigenmodes of a generalized dendrimer with functionality 
$f_c$ for the central core  and  functionality $f$ for the other
junctions belong to two general classes: for the eigenmodes of class (i)
the core is mobile while for those of class (ii) it is immobile.

The non-degenerate eigenvalues $\lambda_r$ corresponding to class (i) can be obtain from
Eqs. (6) and (7) of  Ref. \cite{gurtov_new}; one has namely 
\begin{equation}
\lambda_r  =  f-2\sqrt{f-1}\cos\phi_r, \label{D1}
\end{equation}
where $\phi_r$ is solution of the equation
\begin{equation}
\sin(g+1)\phi_r = \frac{f-f_c-1}{\sqrt{f-1}}\sin g\phi_r. \label{D2} 
\end{equation}
Setting $x_r=\cos\phi_r$ these relations are equivalent to
\begin{equation}
\lambda_r  =  f-2x_r\sqrt{f-1} \label{D3}
\end{equation}
and to
\begin{equation}
U_{g}(x_r) = \frac{f-f_c-1}{\sqrt{f-1}}U_{g-1}(x_r), \label{D4}
\end{equation}
which has $g$ solutions $x_r$.
In the case $$(g+1)/g\leq |f-f_c-1|/\sqrt{f-1},$$ however, Eq. (\ref{D2}) has only 
$g-1$ solutions for real valued $\phi_r$. Then there appears one more eigenvalue 
$\Lambda$, which for $(f-f_c-1)>0$ is given by  Eq. (8) of 
Ref. \cite{gurtov_new}
\begin{equation}
\Lambda=f-2\sqrt{f-1}\cosh\psi, \label{D5}
\end{equation}
corresponding to $x_r=\cosh\psi=\cos(i\psi)>1$. Using Eq. (\ref{text_9}) with 
the imaginary angle $i\psi$ transforms Eq. (\ref{D4}) into
\begin{equation}
\sinh(g+1)\psi = \frac{f-f_c-1}{\sqrt{f-1}}\sinh g\psi, \label{D6} 
\end{equation}
as given in Eq. (9) of Ref. \cite{gurtov_new}. In the case that \\ 
$(f-f_c-1)$ is negative, $\Lambda$ is given by  Eq. (10) of 
Ref. \cite{gurtov_new}
\begin{equation}
\Lambda=f+2\sqrt{f-1}\cosh\psi, \label{D7}
\end{equation}
corresponding to $x_r=-\cosh\psi=\cos(\pi+i\psi)<-1$. 
Using Eq. (\ref{text_9}) with 
the complex angle $(\pi+i\psi)$ transforms Eq. (\ref{D4}) into
\begin{equation}
\sinh(g+1)\psi = -\frac{f-f_c-1}{\sqrt{f-1}}\sinh g\psi, \label{D8} 
\end{equation}
as given in Eq. (11) of Ref. \cite{gurtov_new}.

The eigenvalues $\lambda_r$ for the modes of class (ii) that are 
described by Eq. (\ref{D1}) for real valued angle $\phi_r$ obey
Eqs. (12) and (14) of Ref. \cite{gurtov_new} with
\begin{equation}
\sin(m+1)\phi_r = \sqrt{f-1}\sin m\phi_r, ~~ \textrm{with}~ m=1,...,g. \label{D9} 
\end{equation}
This can be rewritten with help of Eq.(\ref{text_9}) as
\begin{equation}
U_m(x_r) = \sqrt{f-1}U_{m-1}(x_r) ~~ \textrm{with}~ m=1,...,g, \label{D10} 
\end{equation}
again setting $x_r=\cos\phi_r$. The degeneracies are 
\begin{equation} \label{D11}
\Delta_m=\left\{\begin{array}{ll}f_c-1 & \mbox{ if } m=g \\
                f_c(f-2)(f-1)^{g-m+1} & \mbox{otherwise.} \end{array}\right.
\end{equation}
This corresponds to Eq. (\ref{degeneracy}) with $f_c=f$. Again, as noted 
in  Ref. \cite{gurtov_new}, in the case that $m+1\leq m\sqrt{f-1}$, 
Eq. (\ref{D9}) does
not give all the eigenvalues of the modes in terms of real valued $\phi_r$. The
remaining eigenvalue is given by Eq. (\ref{D5}), and it corresponds to $x_r=\cosh\psi>1$, where now $\Psi$ is solution to
Eq. (15) of Ref. \cite{gurtov_new}
\begin{equation}
\sinh(m+1)\psi = \sqrt{f-1}\sinh m\psi \label{D12}.
\end{equation}
This can again be incorporated into Eq.(\ref{D10}), by making the  identification  
$x_r=\cosh\psi=\cos i\psi$.


\begin{thebibliography}{30}
\bibitem{dushek} K. Du$\check{s}$ek and M. Du$\check{s}$kov$\acute{a}$-Smr$\check{c}$kova, {\it Macromolecules} {\bf 36}, 2915 (2003).
\bibitem{gasilova} E. Gasilova, L. Benyahia, D. Durand and T. Nicolai, {\it Macromolecules} {\bf 35}, 141 (2002).
\bibitem{sukumar} V.S. Sukumar and S. T. Lopina, {\it Macromolecules} {\bf 35}, 10189 (2002).
\bibitem{prochazka} F. Prochazka, T. Nicolai and  D. Durand, {\it Macromolecules} {\bf 33}, 1703 (2000).
\bibitem{burchard} W. Burchard, {\it Adv. Polym. Sci.} {\bf 43}, 120 (1999).
\bibitem{erwan2001} N. Erwan, T. Nicolai and D. Durand, {\it Macromolecules} {\bf 34}, 5205 (2001).
\bibitem{vicsek} A. Blumen, A. Jurjiu, T. Koslowski and C. von Ferber, {\it Phys. Rev. E} {\bf 67}, 061103 (2003).
\bibitem{vamvakaki} M. Vamvakaki and C.S. Patrickios, {\it Chem. Mater.} {\bf 14}, 1630 (2002).
\bibitem{doi_edwards} M. Doi and S. R. Edwards, {\it The Theory of Polymer Dynamics}, Clarendon: Oxford, U.K., (1986).
\bibitem{rama} P. Biswas, R. Kant and A. Blumen, {\it Macromolec. Theory Simul.} {\bf 9}, 56 (2000).
\bibitem{sommer_blumen} J.-U. Sommer and A. Blumen, {\it J. Phys. A} {\bf 28}, 6669 (1995).
\bibitem{gurtovenko_blumen} A. A. Gurtovenko and A. Blumen, {\it Adv. Polymer Sci.}, in press.
\bibitem{sgb1} C. Satmarel, A.A. Gurtovenko and A. Blumen, {\it Macromolecules} {\bf 36}, 486 (2003).
\bibitem{sgb2} C. Satmarel, A.A. Gurtovenko and A. Blumen, {\it Macromol. Theory Simul.} {\bf 13}, 487 (2004).
\bibitem{kloczkowski} A. Kloczkowski, J.E. Mark and H.L.Frisch, {\it Macromolecules} {\bf 23}, 3481 (1990).
\bibitem{grassley} W.W. Grassley, {\it Macromolecules} {\bf 13}, 372 (1980).
\bibitem{gurt02} A.A. Gurtovenko and A. Blumen, {\it Macromolecules} {\bf 35}, 3288 (2002).
\bibitem{wedges} A.A. Gurtovenko, Ya.Ya. Gotlib and A. Blumen, {\it Macromolecules} {\bf 35}, 7481 (2002).
\bibitem{cai} C. Cai and Z.Y. Chen, {\it Macromolecules} {\bf 30}, 5104 (1997).
\bibitem{Abramovitz} M. Abramowitz and A. Stegun, {\it Handbook of Mathematical Functions}, Dover Publications, Inc, New York, (1972).
\bibitem{rammal} R. Rammal, {\it J. Phys.} (France), {\bf 45}, 191 (1984).
\bibitem{zimm} B. H. Zimm and R. H. Kilb, {\it J. Polymer Science} {\bf 37}, 19 (1959).
\bibitem{gurtov_new} A.A. Gurtovenko, D. A. Markelov, Ya.Ya. Gotlib and A. Blumen, {\it J. Chem. Phys.} {\bf 119}, 7579 (2003).
\bibitem{gurtov2} A.A. Gurtovenko and Yu.Ya. Gotlib, {\it Macromolecules} {\bf 31}, 5756 (1998).
\bibitem{ferla} R. La Ferla, {\it J. Chem. Phys.} {\bf 106}, 688 (1997).
\bibitem{markelov_new} Yu.Ya. Gotlib and D. A. Markelov, {\it Polym. Sci. A} {\bf 44}, 1341 (2002).
\bibitem{jasch} F. Jasch, C. von Ferber and A. Blumen, {\it Phys. Rev. E} {\bf 68}, 051106 (2003).
\bibitem{ferber} C. von Ferber and A. Blumen, {\it J. Chem. Phys} {\bf 116}, 8616 (2002).
\bibitem{jfb} F. Jasch, C. von Ferber and A. Blumen, {\it Phys. Rev. E} {\bf 17}, 016112 (2004).









\end{thebibliography}
\end{document}